\documentclass[preprint2]{aastex}

\shorttitle{PolarBase}
\shortauthors{Petit et al.}

\begin{document}


\title{PolarBase: a data base of high resolution spectropolarimetric stellar observations}


\author{P. Petit, T. Louge, S. Th\'eado and F. Paletou}
\affil{Institut de Recherche en Astrophysique et Plan\'etologie, Universit\'e de Toulouse \& CNRS, F-31400 Toulouse, France}
\author{N. Manset}
\affil{Canada-France-Hawaii Telescope Corporation, 65-1238 Mamalahoa Hwy Kamuela HI 96743, USA}
\author{J. Morin}
\affil{LUPM, CNRS \& Universit\'e Montpellier II, F-34095 Montpellier, France}
\author{S.C. Marsden}
\affil{Computational Engineering and Science Research Centre, University of Southern Queensland, Toowoomba, 4350, Australia}
\and
\author{S.V. Jeffers}
\affil{Institut f\"ur Astrophysik, Georg-August-Universit\"at G\"ottingen, Friedrich-Hund-Platz 1, 37077 G\"ottingen, Germany}


\begin{abstract}
PolarBase is an evolving data base that contains all stellar data collected with the ESPaDOnS and NARVAL high-resolution spectropolarimeters, in their reduced form, as soon as they become public. As of early 2014, observations of 2,000 stellar objects throughout the Hertzsprung-Russell diagram are available. Intensity spectra are available for all targets, and the majority of the observations also include simultaneous spectra in circular or linear polarization, with the majority of the polarimetric measurements being performed only in circularly polarized light (Stokes V). Observations are associated with a cross-correlation pseudo-line profile in all available Stokes parameters, greatly increasing the detectability of weak polarized signatures. Stokes V signatures are detected for more than 300 stars of all masses and evolutionary stages, and linear polarization is detected in 35 targets. The detection rate in Stokes V is found to be anti-correlated with the stellar effective temperature. This unique set of Zeeman detections offers the first opportunity to run homogeneous magnetometry studies throughout the H-R diagram. The web interface of PolarBase is available at http://polarbase.irap.omp.eu.    
\end{abstract}





\section{Introduction: the NARVAL and ESPaDOnS spectropolarimeters}

\begin{figure}
\includegraphics[width=\columnwidth]{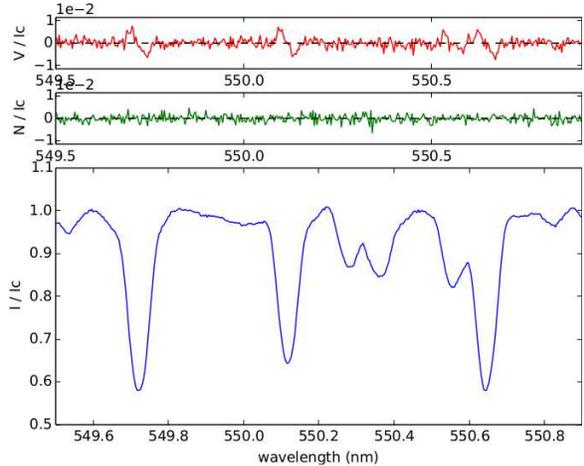}
\caption{Extract of a normalized spectrum of II~Peg, observed with ESPaDOnS on 2005 July 14, around three Fe \textsc{I} spectral lines with high Land\'e factors. The Stokes V and I spectra are illustrated in the top and bottom panels, respectively. Three Zeeman signatures are clearly visible in the Stokes V spectrum. The middle panel displays one of the two ``Null" spectra. Note that Ic stands for the continuum intensity. (after \citealt{petit06})}
\label{fig:iipeg}
\end{figure}

The ESPaDOnS spectropolarimeter \citep{donati06b} has been operational at the Canada-France-Hawaii Telescope (CFHT, Mauna Kea Observatory) since early 2005. NARVAL \citep{auriere03} is a strict twin of ESPaDOnS, available at T\'elescope Bernard Lyot (TBL, Pic du Midi Observatory) since late 2006. The development of these two high-resolution optical spectropolarimeters was primarily motivated by their capabilities of detecting and characterizing stellar magnetic fields through the Zeeman effect.  This science objective imposed strong constraints on the instrumental design, which has to combine a polarimeter and a spectrograph.

Their polarimetric module performs measurements in circular (Stokes V) or linear (Stokes Q \& U) polarization, and is installed at the Cassegrain focus of the telescope to avoid any oblique reflection prior to the polarimetric analysis. Once split into two orthogonal polarization states, the light emerging from the polarimeter is sent to the spectrograph through a fiber link. As will be discussed in Section \ref{sec:lsd}, the polarimetric sensitivity of modern instruments is greatly enhanced by their capability of repeating the measurement of Zeeman signatures over a large number of spectral lines. To cover a wide spectral window in a single exposure, ESPaDOnS and NARVAL incorporate a cross-dispersed echelle spectrograph, offering coverage of the entire optical domain (370~nm to 1,000~nm), with limited gaps of a few nanometers in the near-infrared portion of the spectrum. Both spectropolarimeters have a spectral resolving power of $\approx 65,000$ in polarimetric mode, as required to resolve narrow spectral lines and therefore recover the detailed shape of polarized Zeeman signatures present in individual spectral features. They also have a higher resolving power of  $\approx 76,000$ when used for classical spectroscopy alone. For a more detailed description of ESPaDOnS and NARVAL, including a comparison of their performances and recent instrumental upgrades, we refer the reader to the very complete paper of \cite{silvester12}.

The polarimetric accuracy, spectral resolution, spectral coverage and global throughput of ESPaDOnS and NARVAL constitute an impressive improvement compared to their predecessors, mainly MuSiCoS at TBL \citep{baudrand92,donati99} and SemelPol at the Anglo-Asutralian Telescope \citep{donati03}. This leap forward has already led to spectacular scientific results involving magnetic field measurements throughout most of the Hertzsprung-Russell diagram (HR diagram from now on). Classes of stars explored by the twin spectropolarimeters include massive stars \citep{donati06a}, Ap stars \citep{auriere07}, Vega-like stars \citep{lignieres09}, Sun-like stars \citep{petit08,marsden13}, M dwarfs \citep{morin08,morin10}, cool giants and supergiants \citep{auriere09,auriere10} and young stars \citep{donati07,alecian08}.

Public ESPaDOnS and NARVAL data are separately made available by their host observatory. Raw and reduced ESPaDOnS data are offered at CADC (http://www2.cadc-ccda.hia-iha.nrc-cnrc.gc.ca) and reduced NARVAL data can be accessed through the TBL Legacy data base (http://tblegacy.bagn.obs- mip.fr). PolarBase is grouping all reduced data together with enhanced query and online plotting facilities. More advanced modeling is also offered in addition to the reduced data, in the form of cross-correlation pseudo-line profiles and magnetic detection flags.

\section{Reduced spectropolarimetric data}
\label{sec:reduction}

The data reduction is automatically performed in real-time at TBL and CFHT, using the Libre-Esprit reduction pipeline specifically developed for both twin spectropolarimeters and implementing the algorithm described by \cite{donati97}. The reduction software performs optimal extraction of the spectrum (in all available Stokes parameters), following the approach of \cite{marsh89} and \cite{horne86}. A first wavelength calibration is achieved using Th-Ar spectra taken during the night of observation, and is refined at the end of the reduction process by using telluric bands superimposed on each stellar spectrum as radial velocity references. Using this strategy, the final radial velocity accuracy is of the order of 20-30 $ms^{-1}$ \citep{moutou07}.  

Libre-Esprit produces spectra with and without continuum normalization. PolarBase offers both types of reduction output, except when one of the modes did was rejected by our quality checks (see Section \ref{sec:targets}), or when one of the reduction modes was not initially computed by the observatory. Note that any continuum polarization is subtracted by default during data reduction, as the optical design of ESPaDOnS and NARVAL is not optimized for continuum polarization measurements.

In polarimetric mode, spectra are produced by the combination of 4 sub-exposures taken with the polarimetric optics rotated at different angles about the optical axis, following the approach of \cite{semel93}. The corresponding Stokes I spectrum is obtained by adding together the 4 sub-exposures (note that Stokes I spectra are also derived from each sub-exposure and made available in PolarBase). Using other combinations of the sub-exposures, it is possible to obtain two ``Null" spectra \citep{donati97} that are not expected to contain any signal. The ``Null" spectra are used to check for spurious polarized signatures (of stellar or instrumental origin). Fig. \ref{fig:iipeg} illustrates, over a small wavelength domain, the Stokes I, V and ``Null" spectra of a normalized ESPaDOnS observation reduced with Libre-Esprit. 

The PolarBase spectra are made available in their native ASCII format, where data are organized in columns. Table \ref{tab:data} lists the content of the ASCII files, depending on the data type. The number of spectral bins is given in the second line of the file, as well as the number of data columns (which depends on the data type). In polarimetric mode, 5 data columns are provided, so that the file contains a total of 6 columns, if we also consider the left column listing the wavelength (expressed in nanometers). In spectroscopic mode, only 2 data columns are listed (plus one for wavelength). ESPaDOnS offers a seldom-used ``star+sky" spectroscopic mode, where the sky light is injected in a dedicated fiber. In PolarBase ASCII files produced for this mode, the intensity column contains the stellar spectrum from which the sky contribution was subtracted.  

For each spectrum, a second ASCII file is also provided that contains meta-data relevant to the observation. The target name, coordinates, time of observation (in date/time format, Heliocentric Julian date or Geocentric Julian date), and signal-to-noise ratio in each spectral order can be found in this complementary file.

\begin{table}
\centering
\caption[]{Format of ASCII PolarBase data files for reduced spectra (``Spec.") and LSD profiles (``LSD"). ``Wav." stands for the wavelength (in nanometers) and ``RV" stands for the radial velocity (in km/s). ``Int." and ``Pol." correspond to intensity and polarized data (normalized or not). ``Null 1" and ``Null 2" are the check parameters described in Sec. \ref{sec:reduction}. ``$\sigma$" stands for the uncertainty on a given type of data.}
\begin{tabular}{ccccc}
\hline
Col. & Spec. & Spec. & LSD & LSD \\
\# & (pol) & (I) & (pol) & (I) \\
\hline
0 & Wav. & Wav. & RV & RV \\
1 & Int. & Int. & Int. & Int. \\
2 & Pol. & $\sigma_{\rm int}$ & $\sigma_{\rm int}$ & $\sigma_{\rm int}$ \\
3 & Null 1 & -- & Pol. & -- \\
4 & Null 2 & -- & $\sigma_{\rm pol}$ & -- \\
5 & $\sigma_{\rm int}$ & -- & Null 1 & -- \\
6 & -- & -- & $\sigma_{\rm null 1}$ & --\\
\hline
\end{tabular}
\label{tab:data}
\end{table}

\section{Least-Squares Deconvolution of spectra}
\label{sec:lsd}

\begin{figure}
\includegraphics[width=\columnwidth]{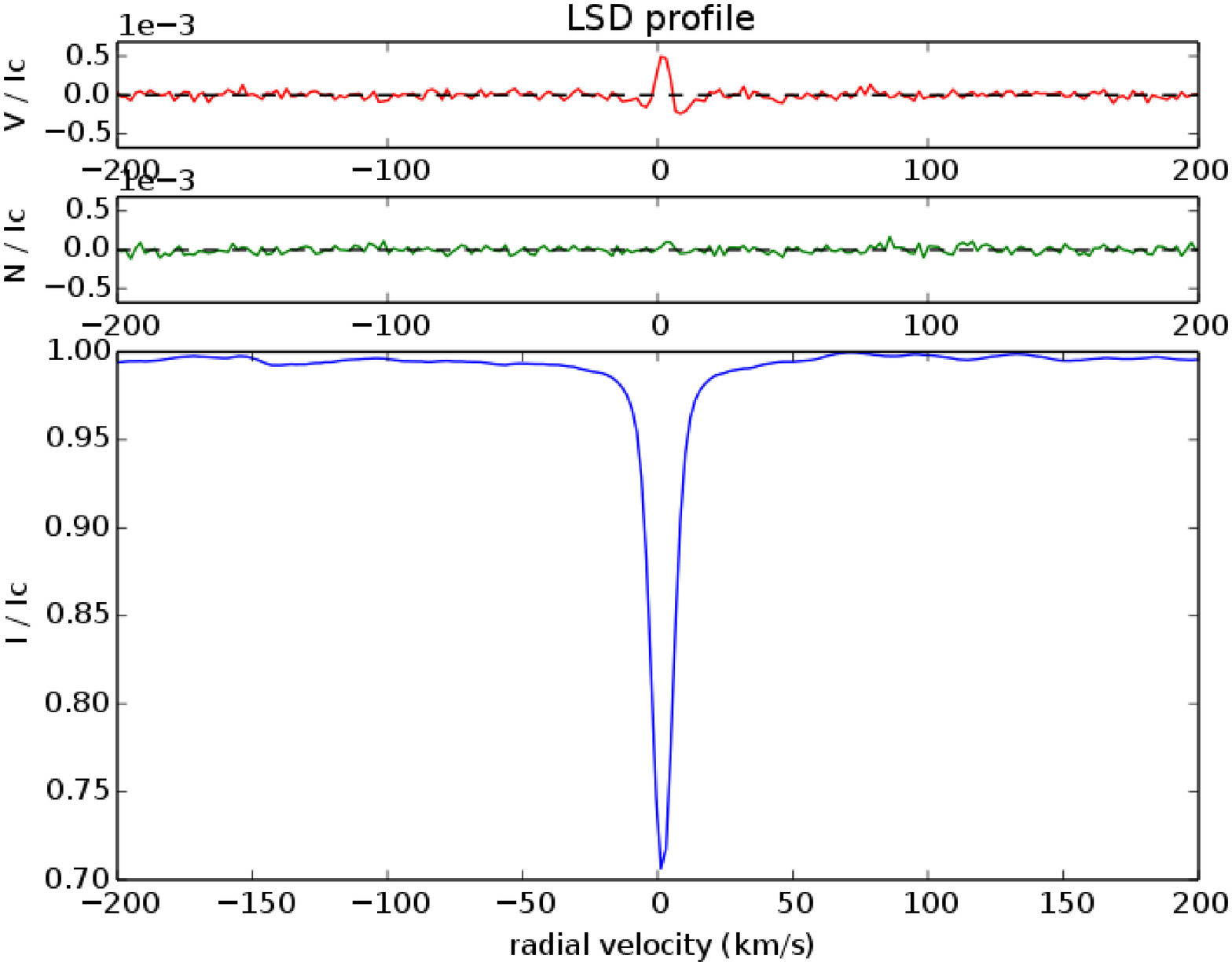}
\caption{Example LSD profile for the G8 dwarf $\xi$~Bootis~A, observed with NARVAL on 2008 February 11. The Stokes V and I cross-correlation profiles are illustrated in the top and bottom panels, respectively. A strong Zeeman signature is detected in the Stokes V line profile. The middle panel illustrates one of the two ``Null" LSD profiles. Ic stands for the continuum intensity. (after \citealt{morgenthaler12}) }
\label{fig:targets}
\end{figure}

\begin{figure}
\includegraphics[width=\columnwidth]{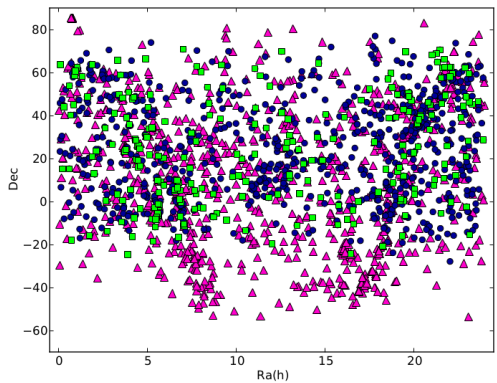}
\caption{PolarBase targets located in a sky map. The right ascension is expressed in hours, and the declination is expressed in degrees. Pink triangles and black circles are objects observed by ESPaDOnS and NARVAL, respectively, while green squares illustrate stars observed by both instruments.}
\label{fig:targets}
\end{figure}

The capability of ESPaDOnS and NARVAL for detecting very weak polarized signatures (most of the time below the noise level of reduced spectra) is improved by (and, most of the time, requires) the simultaneous extraction of the polarimetric signal from all available photospheric lines.

Each normalized PolarBase spectrum was processed with the Least-Squares Deconvolution method (LSD thereafter, \citealt{donati97,kochukhov10}), to extract a mean pseudo-line profile in all available Stokes parameters, using here the implementation of \cite{donati97}. Here, free parameters in LSD models are optimized by following a procedure similar to the one employed by \cite{marsden13}, and the resulting scale factors are listed in Tables \ref{tab:scale1} and \ref{tab:scale2}. This multi-line approach requires the selection of a list of spectral lines that matches the photospheric properties of the target. The line-list used to compute the LSD profile is created from a query of the VALD atomic data base \citep{kupka00}, with lines in the wings of Balmer and Helium lines removed from the list. In the current operation of PolarBase, we rely on the SIMBAD data base to provide us with an approximate spectral type of the target, from which we select a line-mask of the closest effective temperature ($T_{\rm eff}$), using the relation between spectral type and temperature given by \cite{gray05}. We crudely assume that the surface gravity of the star is a main-sequence surface gravity (at the $T_{\rm eff}$ of the target) and also assume a solar metallicity. When the target is a SB2 (or any more complex stellar association), the line-mask is chosen to match the effective temperature of the primary component. 

Considering this series of crude approximations, we recommend to limit the use the PolarBase LSD profiles as a preliminary help to check for the presence of a polarized signature. It is our experience that an incorrect line-mask will not produce a spurious polarized signature in a non-magnetic target. However, an incorrect line-list will reduce the amplitude of the polarized signature in the LSD profile, so that some weak Zeeman signatures may stay below the noise level, whereas a better line list might have revealed their presence.

PolarBase LSD line profiles are downloadable in ASCII format, following the column distribution described in Table \ref{tab:data}. As for reduced spectra, basic information about the LSD model (e.g. S/N of the LSD profile) are listed in a separate ASCII file. The second file also gives a probability of signal detection, following the prescription of \cite{donati92}. Detection probabilities are computed within the line profile and in the neighbouring continuum, both for the relevant Stokes parameter and for its associated ``Null" line profile. A flag is given that classifies the detection probability as a ``definite", ``marginal" or ``no" signal detection, following the conventions detailed by \cite{donati97}. 

\section{Data base}

\begin{table}
\centering
\caption[]{Table of scale factors used to generate the LSD profiles (spectral types O to F).}
\begin{tabular}{cccc}
\hline
Sp. Type & $d_0$ & $g_0$ & $\lambda_0$ \\
\hline
O4 & 0.1693 & 1.110 & 541.3416 \\
O5 & 0.1693 & 1.116 & 549.3795 \\
O6 & 0.1671 & 1.108 & 537.3378 \\
O7 & 0.1693 & 1.109 & 536.1185 \\
O8 & 0.1933 & 1.141 & 535.6516 \\
O9 & 0.1959 & 1.140 & 510.0581 \\
\hline
B0 & 0.2307 & 1.240 & 510.5091 \\
B1 & 0.2361 & 1.256 & 533.1491 \\
B2 & 0.2377 & 1.257 & 573.0493 \\
B3 & 0.2593 & 1.231 & 535.3934 \\
B4 & 0.2874 & 1.185 & 545.0606 \\
B5 & 0.2687 & 1.191 & 612.7151 \\
B6 & 0.3266 & 1.138 & 580.1827 \\
B7 & 0.3556 & 1.119 & 571.3179 \\
B8 & 0.3325 & 1.163 & 711.4965 \\
B9 & 0.4090 & 1.166 & 510.6888 \\
\hline
A0 & 0.4143 & 1.209 & 501.7136 \\
A1 & 0.4219 & 1.198 & 542.4850 \\
A2 & 0.4219 & 1.198 & 543.4444 \\
A3 & 0.4303 & 1.201 & 545.0059 \\
A4 & 0.4303 & 1.201 & 545.0059 \\
A5 & 0.4524 & 1.201 & 556.1013 \\
A6 & 0.4657 & 1.208 & 550.2700 \\
A7 & 0.4625 & 1.209 & 557.1888 \\
A8 & 0.4625 & 1.209 & 557.1888 \\
A9 & 0.4688 & 1.206 & 555.6388 \\
\hline
F0 & 0.4688 & 1.206 & 555.6388 \\
F1 & 0.4877 & 1.212 & 556.8212 \\
F2 & 0.4877 & 1.212 & 556.8212 \\
F3 & 0.4877 & 1.212 & 574.2867 \\
F4 & 0.4812 & 1.210 & 581.6153 \\
F5 & 0.4812 & 1.210 & 581.6153 \\
F6 & 0.5053 & 1.213 & 582.9365 \\
F7 & 0.5053 & 1.213 & 582.9365 \\
F8 & 0.5294 & 1.212 & 577.4088 \\
F9 & 0.5225 & 1.211 & 579.8654 \\
\hline
\end{tabular}
\label{tab:scale1}
\end{table}

\begin{table}
\centering
\caption[]{Same as Table \ref{tab:scale1} for spectral types G to M.}
\begin{tabular}{cccc}
\hline
Sp. Type & $d_0$ & $g_0$ & $\lambda_0$ \\
\hline
G0 & 0.5294 & 1.211 & 577.3031 \\
G1 & 0.5225 & 1.210 & 584.1816 \\
G2 & 0.5225 & 1.210 & 597.6029 \\
G3 & 0.5225 & 1.208 & 605.4862 \\
G4 & 0.5225 & 1.208 & 605.4862 \\
G5 & 0.5363 & 1.210 & 589.6192 \\
G6 & 0.5363 & 1.210 & 589.6192 \\
G7 & 0.5486 & 1.207 & 601.1477 \\
G8 & 0.5486 & 1.207 & 601.1477 \\
G9 & 0.5554 & 1.208 & 615.4197 \\
\hline
K0 & 0.5554 & 1.208 & 615.4197 \\
K1 & 0.5550 & 1.210 & 603.7351 \\
K2 & 0.5625 & 1.214 & 612.8669 \\
K3 & 0.5625 & 1.214 & 612.8669 \\
K4 & 0.6887 & 1.206 & 629.8284 \\
K5 & 0.6887 & 1.206 & 629.8284 \\
K6 & 0.6887 & 1.206 & 629.8284 \\
K7 & 0.6887 & 1.206 & 629.8284 \\
K8 & 0.6751 & 1.221 & 665.0525 \\
K9 & 0.6751 & 1.221 & 665.0525 \\
\hline
M0 & 0.6751 & 1.221 & 665.0525 \\
M1 & 0.6751 & 1.221 & 665.0525 \\
M2 & 0.6188 & 1.245 & 672.9030 \\
M3 & 0.6188 & 1.245 & 672.9030 \\
M4 & 0.6188 & 1.245 & 672.9030 \\
M5 & 0.6188 & 1.245 & 672.9030 \\
M6 & 0.6188 & 1.245 & 672.9030 \\
M7 & 0.6188 & 1.245 & 672.9030 \\
M8 & 0.6188 & 1.245 & 672.9030 \\
\hline
\end{tabular}
\label{tab:scale2}
\end{table}

The technical architecture of PolarBase is based upon PostgreSQL9.0 and Python scripts. The full software is divided into three different parts:
\begin{itemize}
\item Data integration software, designed to feed the database with new sets of data released each semester by TBL and CFHT. The corresponding algorithms are implemented as object-designed Python scripts. 
\item The query software (developed in Python) is used to access the database and manipulate data through the user web interface. 
\item User web interface.
\end{itemize}

The integration and query software make use of the SIMBAD data base to collect basic stellar parameters of the targets and store these meta-data into a specific section of the database. Metadata missing in SIMBAD are attributed the ``-999" value in PolarBase. Regular SIMBAD checks are performed to update the list of available metadata. There are two types of data available in PolarBase, with reduced spectra (normalized or not) and LSD line profiles. Each one is treated through an independent integration pipeline and consistency check, looking for corrupted data (wrong format or inconsistent quantities) and evaluating the data quality (e.g. through thresholds in the signal-to-noise ratio) to make sure that the content available to the user is of high quality. 

\section{Web interface and virtual observatory}

The user interface uses AJAX with Rialto, Jquery and Dygraphs frameworks to give the user a state-of-the-art browsing experience. All data available can be accessed through this interface (http://polarbase.irap.omp.eu). The query criteria are divided into different groups:
\begin{itemize}
\item Object characteristics (name, magnitude, spectral type...)
\item Parameters related to a specific observation (airmass, date, observing mode, stokes parameter...)
\item Data quality and LSD results (signal-to-noise ratio, detection of a polarimetric signal)
\end{itemize}

When the database is queried for a specific object name (given alone or along with other parameters), we first check whether the specified name is found in the data base. If not, then a SIMBAD query is used to cross-match the name given by the user, SIMBAD results and PolarBase content. The database then returns a list of observation blocks matching requested criteria. The blocks contain spectra (normalized and not normalized) and LSD output that can all be plotted, zoomed in and out and downloaded. 

In addition to the ASCII files accessible through its web interface, Polarbase offers FITS format files through its VO portal, with VO services directly derived from the set of services available for the TBLegacy archive (http://tblegacy.bagn.obs-mip.fr).

\section{Stellar targets}
\label{sec:targets}

\begin{figure}
\includegraphics[width=\columnwidth]{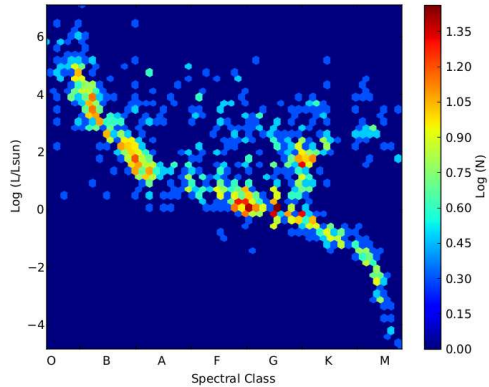}
\caption{PolarBase targets located in the Hertzsprung-Russell diagram. The N quantity in the color bar stands for the number of objects in each bin. Stellar luminosities are derived from apparent V magnitudes and parallax measurements listed in SIMBAD, as well as bolometric corrections taken from \cite{gray05}.}
\label{fig:hr}
\end{figure}

\begin{figure}
\includegraphics[width=\columnwidth]{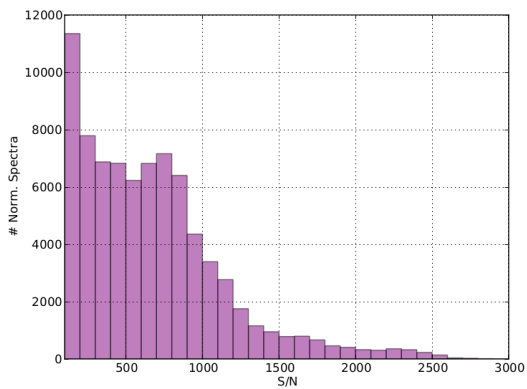}
\caption{Number of normalized PolarBase spectra, as a function of their peak signal-to-noise ratio per 1.8 km.s$^{-1}$ velocity bin in Stokes I (before LSD processing).}
\label{fig:sn}
\end{figure}

\begin{figure}
\includegraphics[width=\columnwidth]{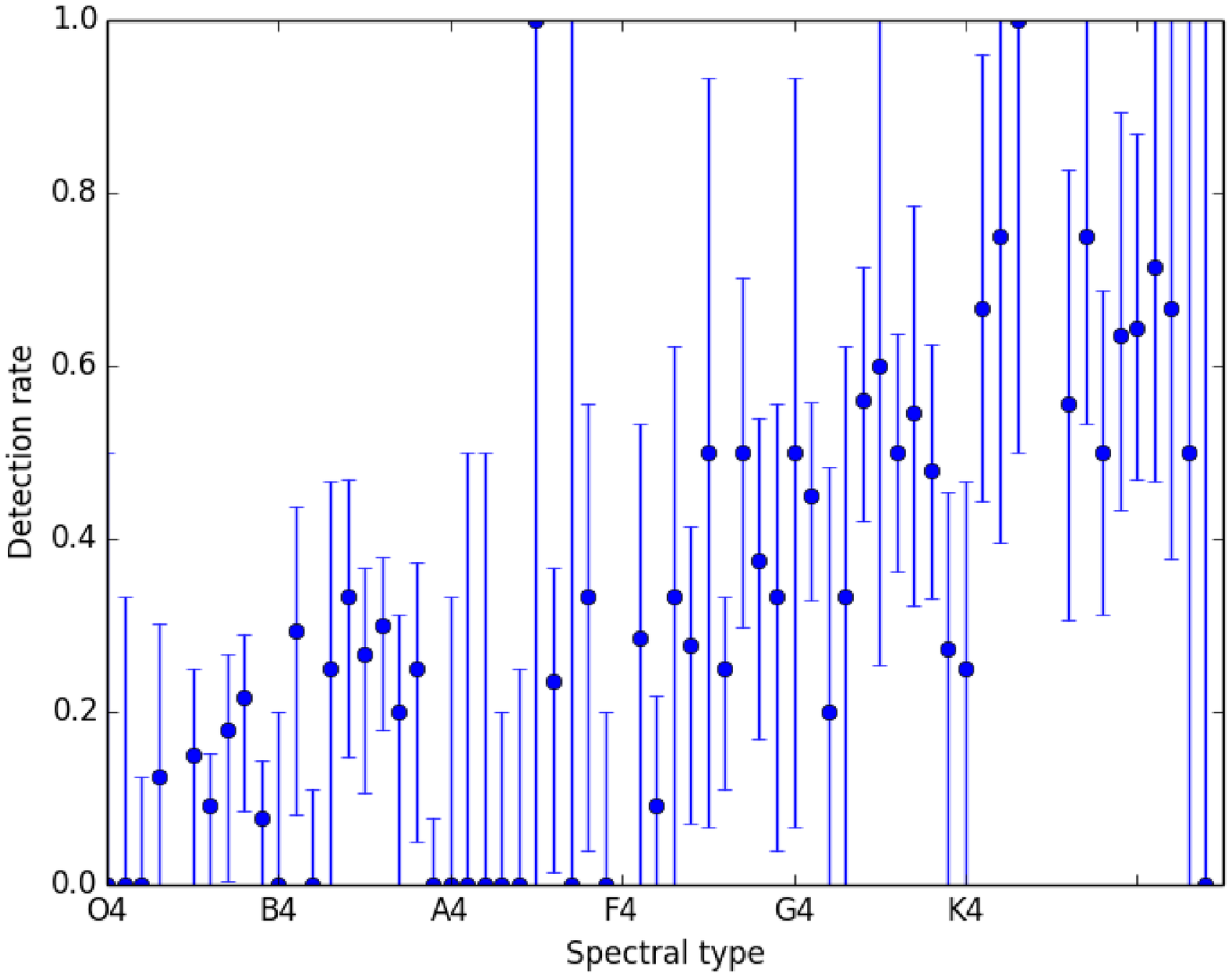}
\caption{Fraction of objects for which Stokes V signatures were detected at least once in LSD line profiles, as a function of spectral type. Anticorrelation is observed between detection rate and effective temperature. The uncertainties are computed from binomial statistics.}
\label{fig:detection}
\end{figure}

Among all objects observed with ESPaDOnS and NARVAL, we have selected for PolarBase stellar targets with a SIMBAD entry. We therefore exclude all observations of solar-system bodies and extra-galactic objects, as well as stellar observations with object names that could not be resolved using the SIMBAD data base. Taking into account all public observations available so far (collected before 2013), we end up with around 1290 stars observed with ESPaDOnS and 961 stars observed with NARVAL, corresponding to a total of 2004 observed targets (247 objects being observed with both instruments).  

The distribution of PolarBase targets over the sky is shown in Fig. \ref{fig:targets}, displaying a denser accumulation of observations in the plane of the Milky Way. Thanks to the location of Mauna Kea observatory at lower latitude, ESPaDOnS offers a better sampling than NARVAL at lower declination (down to $-50^\circ$, versus a minimum declination of $-20^\circ$ with NARVAL). Due to pointing limitations of TBL at high declinations (with a maximum accessible declination of around $75^\circ$), ESPaDOnS also gives easier access to stars closer to the northern celestial pole (maximum observed declination of $85^\circ$ in PolarBase).

PolarBase stars sample a wide range of spectral types, from O4 to M9 (Fig. \ref{fig:hr}), with a 60\% majority of the targets being cooler than F5. In the HR diagram, PolarBase targets are mainly distributed over the main sequence, with a relative lack of targets of A specral type. A number of stars are also located at higher luminosities, including both pre-main-sequence targets and evolved objects. We note a dearth of stars in the lower-left quadrant of the diagram, showing the scarcity of observing programs dedicated to compact objects. A 60\% majority of stars hotter than F5 were observed with ESPaDOnS, owing to the contribution of the MiMeS survey \citep{wade13}, while the observation of targets cooler than F5 was more evenly distributed between both instruments.    

\section{Characteristics of observed spectra}

The content of PolarBase was cleaned from low quality spectra, using simple filters to automatically remove problematic observations. In particular, we discarded 5,755 spectra for which the pointing coordinates differed from the object coordinates by more than $1^\circ$ (in right ascension or declination). We also removed from the data base 10,634 spectra with peak S/N lower than 100 (most of them concerning cool stars), and  1,328 observations with peak S/N greater than 3,000 (since they are likely affected by partial CCD saturation). The search for tiny Zeeman signatures requires high S/N ratios, so that low S/N spectra are a minority in PolarBase, with 58\% of available spectra having a peak S/N above 500 and 18\% of them above 1,000 before LSD processing (see Fig. \ref{fig:sn}).

Most PolarBase objects are variable and were therefore repeatedly observed with ESPaDOnS or NARVAL. As of early 2014, the total number of independent observations is equal to $\approx 90,000$. Since most observations are offered with and without continuum normalization, the total number of spectra in PolarBase is about twice this amount. Queries for observations of a given object at a specific date are possible with the web interface of PolarBase.

\section{Polarimetric signatures}

A 64\% majority of PolarBase targets where observed in the Stokes V polarimetric mode, and 8\% of the observed stars benefit from linear polarization measurements (most of them being intermediate-mass stars). Among all stars observed at least once in Stokes V, 24\% provided us with at least one definite detection in the LSD line profile, leading to a total of 307 objects with polarimetric signatures that are likely attributable to a photospheric magnetic field.

As a simple example of data mining that can be easily performed with PolarBase, we show in Fig. \ref{fig:detection} the fraction of stars with at least one signal detection in Stokes V, as a function of spectral type. An anti-correlation between field detection and effective temperature is clearly observed in this plot. The lowest detection rate is obtained for O stars, with a 5\% detection ration consistent with findings from the MiMeS survey \citep{wade13}. Much higher detection rates of around 50\% are obtained for K and M spectral types. Interestingly, detection rates from late F to early K matches quite well the 40\% detection rate reported by \cite{marsden13} from the Bcool catalogue of main-sequence stars, although the present plot also incorporates pre-main-sequence stars and cool evolved stars at these spectral types. Our detection rate of about 20\% for late-B and early-A stars is somewhat higher than previous estimates by \cite{wolff68} and \cite{vogt98}, although our uncertainties are relatively large. It is also interesting to note the lack of any magnetic detection for late A stars. 

We must however stress that several biases may affect these detection rates in one direction or the other. First, we emphasize again that the line-masks used for our LSD models are approximate masks chosen from the SIMBAD spectral type of the target. The effect of an approximate line-mask is expected to be more pronounced for O stars, for which the scarcity of available lines requires a more careful selection of spectral features to incorporate in the LSD model. Detection rates reported here may therefore give an underestimated value (\citealt{wade13} however propose 7\%, in close agreement with our value). On the other hand, many observing programs carried out with ESPaDOnS or NARVAL concentrated on targets specifically selected from the assumption that they may host a detectable field, as suspected from e.g. high chromospheric (or coronal) emission in the case of cool stars. This is particularly true for M dwarfs that were selected from their high activity \citep{morin08,morin10}, so that the PolarBase sample of M stars may lead to an overestimate of field detection rates in the coolest part of the HR diagram. The lack of magnetic detections in A stars cooler than A3 is likely biased by the low number of NARVAL and ESpaDOnS programs dedicated to Ap stars, but it may also confirm a drop of the prevalence of the Ap phenomenon in stars cooler than A4 \citep{power07}.    

\section{Future science and VO developments}

The data content of PolarBase is meant to progressively evolve. The main reason is the growing amount of public data available for inclusion in the data base, with new observations released every semester by the host observatories of ESPaDOnS and NARVAL. 

PolarBase is also evolving to offer an increasing number of systematic measurements derived from the spectra and LSD profiles. Future quantities that will progressively enrich spectral analysis include the S-index (a classical quantity evaluating Ca II H \& K core emission) and other similar chromospheric emission indices computed for $H_\alpha$ and for the Ca II infrared triplet (following the work of \citealt{marsden13}).  

LSD models will progressively gain better accuracy with the use of refined line-masks. This step will however require a more precise determination of stellar fundamental parameters, through spectral classification tools applied to PolarBase spectra. Quantities derived from the improved LSD models will then include radial velocity and $v.\sin (i)$ measurements (from Stokes I LSD profiles), and longitudinal field measurements using the first moment of Stokes V line profiles \citep{rees79}. 

A very important aspect of our work is to make sure that PolarBase always matches up to the latest revisions of the IVOA (International Virtual Observatory Alliance) formats and protocols (description of flux provided in the data, BAND query parameter and other updates) of both SSAP (Simple Spectral Access Protocol Version 1.1, IVOA Recommendation of 10 February 2012) and SDM (IVOA Spectral Data Model Version 2.0, IVOA proposed Recommendation of 25 April 2013, under review). A first improvement will be that Polarbase will not only offer SSAP access but also a Cone Search service. A second objective is to share LSD models through the VO, which will require a VO compliant description of their content. Extension of the VO capabilities of Polarbase to the use of the Parameter Description Language use (PDL Version 1.0, IVOA Proposed Recommendation of 14 November 2013, document still under review) will also be investigated to ensure optimal interoperability with other VO providers. 

\acknowledgments

This research has made use of the SIMBAD database operated at CDS, Strasbourg, France, and of NASA's Astrophysics Data System. PolarBase data are a compilation of a large number of observing runs, and we are deeply grateful to all the PIs for their talented work in acquiring this exceptional library of spectra. We are grateful to the referee, G.A. Wade, for constructive comments and very useful bug reports. This work has only been possible due to the brilliance of the late Meir Semel, from Observatoire de Paris-Meudon. Sadly Meir passed away during 2012, but his memory and influence on spectropolarimetry will live on. 



{\it Facilities:} \facility{CFHT}, \facility{TBL}

\end{document}